\documentclass[fleqn,usenatbib]{mnras}

\usepackage{newtxtext,newtxmath}

\usepackage[T1]{fontenc}

\DeclareRobustCommand{\VAN}[3]{#2}
\let\VANthebibliography\thebibliography
\def\thebibliography{\DeclareRobustCommand{\VAN}[3]{##3}\VANthebibliography}

\usepackage{graphicx}	
\usepackage{amsmath}	
\usepackage[switch]{lineno}

\title[MAD in FR I radio galaxies]{Magnetically arrested disks in FR I radio galaxies}

\author[He et al.]{
Han He,$^{1}$
Bei You,$^{1}$\thanks{youbei@whu.edu.cn}
Ning Jiang,$^{2,3}$
Xinwu Cao,$^{4}$
Jingfu Hu,$^{5}$
Zhenfeng Sheng,$^{6,2,3}$
Su Yao$^{7}$
and Bozena Czerny$^{8}$
\\
$^{1}$Department of Astronomy, School of Physics and Technology, Wuhan University, Wuhan 430072, People’s Republic of China\\
$^{2}$Key laboratory for Research in Galaxies and Cosmology, Department of Astronomy, University of Science and Technology of China, Chinese Academy of Sciences, Hefei, Anhui 230026, China\\
$^{3}$School of Astronomy and Space Sciences, University of Science and Technology of China, Hefei, Anhui 230026, China\\
$^{4}$Institute for Astronomy, School of Physics, Zhejiang University, 866 Yuhangtang Rd, Hangzhou, 310058, China\\
$^{5}$School of Physics and Electronic Information Engineering, Qinghai Normal University, Xining, 810000, China\\
$^{6}$Institute of Deep Space Sciences, Deep Space Exploration Laboratory, Hefei 230026, China\\
$^{7}$National Astronomical Observatories, Chinese Academy of Sciences, 20A Datun Road, Beijing 100101, China\\
$^{8}$Center for Theoretical Physics, Polish Academy of Sciences, Al. Lotnikow 32/46, 02-668 Warsaw, Poland
}

\date{Accepted XXX. Received YYY; in original form ZZZ}

\pubyear{2015}
\let\oldequation\equation
\let\oldendequation\endequation
\renewenvironment{equation}{\linenomathNonumbers\oldequation}{\oldendequation\endlinenomath}

\begin{document}
\label{firstpage}
\pagerange{\pageref{firstpage}--\pageref{lastpage}}
\maketitle
\begin{abstract}
A sample of 17 FR I radio galaxies constructed from the 3CR catalog, which is characterized by edge-darkened radio structures, is studied.
The optical core luminosities derived from Hubble Space Telescope observation are used to estimate the Eddington ratios which are found to be below $10^{-3.4}$ for this sample. This is supported by the Baldwin–Phillips-Terlevich optical diagnostic diagrams derived with the spectroscopic observation of Telescopio Nazionale Galileo, suggesting that these sources are of low ionization nuclear Emission-line Regions (LINERs). It implies that the accretion in these FR I sources can be modeled as advection-dominated accretion flows (ADAFs). Given the low accretion rate, the predicted jet power with a fast-spinning black hole (BH) $a=0.95$ in the Blandford-Znajek mechanics is lower than the estimated one for almost all the sources in our sample. Such powerful jets indicate the presence of magnetically arrested disks (MAD) in the inner region of the ADAF, in the sense that the magnetic fields in the inner accretion zone are strong. Moreover, we show that, even in the MAD scenario, the BH spins in the sample are most likely moderate and/or fast with $a\gtrsim0.5$.
\end{abstract}




\section{Introduction}

Radio galaxies are a class of radio-loud active galactic nuclei, having jets that tend to be oriented near the plane of the sky, giving us a view of their radio structure from pc to Mpc scales \citep{urry1995}. They are categorized into Fanaroff-Riley type I and type II galaxies (FR I and FR II; \citealt{fanaroff1974}), according to their morphology, with the former one being of edge-darkened radio structure while the latter one being of edge-brightened radio structure. In past studies, the host galaxies of FR II sources are more luminous, in general, than the hosts of FR I radio galaxies in optical R-band \citep{ghisellini2001}. However, recent work shows a big overlap between the host galaxies of FR I and FR II sources in radio luminosity and optical luminosity plane, by analysing a sample of 5805 radio galaxies and found the majority of low radio luminosity FR IIs are consistent with most FR Is in optical $K_s$-band, and they also found a clear dichotomy only between the brightest FR II and FR I sources \citep{mingo2019}. What causes the difference between the two types of radio galaxies, in terms of both the radio morphology and the luminosities, has been a long-standing problem. It could be due to either the interaction of the jet with the ambient medium, or the intrinsic properties of accretion flow and jet formation processes \citep{gopal-Krishna1988,bicknell1995,baum1995,zirbel1995,ghisellini2001,cao2004,wu2011,best2012,li2014,kino2021,perucho2022}. 

In \cite{ghisellini2001}, the mass of the central black holes (BHs) in a radio galaxy is estimated from the absolute optical R–band magnitude of the host galaxy, and the jet power is estimated from the radio luminosity at 151 MHz \citep{willott1999} in approximation, although the radio luminosity also depends on the environment and the age of the sources. Thus, the division of FR I/II in \cite{ledlow1996} can be depicted in the plane of the jet power and the black hole mass as well. \cite{ghisellini2001} proposed that the ratio of jet power to the black hole mass can be used to classify FR I and FR II radio galaxies, in the sense that the differences in accretion modes are responsible for the differences between FR I and FR II. In this scenario, the accretion mode of FR I radio galaxies belongs to the advection-dominated accretion flow (ADAF; \citealt{narayan1995}) branch as the Eddington-scaled accretion rates (hereafter accretion rate for short) of FR I galaxies are likely smaller than 0.01, while the accretion mode of FR II radio galaxies is a standard disk described by Shakura \& Sunyaev disk (SSD; \citealt{shakura1973}).
However, with the work for accretion mode, a number of studies showed that both FR IIs and FR Is can be produced from SSD or ADAF \citep{cao2004,hardcastle2018,hu2018,mingo2022}. Recently, \cite{mingo2022} demonstrated that low-luminosity FR II sources, with the radio luminosity at 150 MHz smaller than $10^{26}~\rm W~Hz^{-1}$, might be low-accretion rate in low-power, and they also found it might be stellar mass, for the sources in low-power, that determines whether they become FR I or low-luminosity FR II sources. 
In the frame of unification schemes, FR I radio galaxies are believed to be BL Lacertae (BL Lac) objects with the relativistic jet aligned to our line of sight, and high-luminosity FR II radio galaxies correspond to misaligned radio quasars. The Eddington ratios $L_{\rm bol}/L_{\rm Edd}$ of BL Lac objects are systematically lower than the ones of quasars, at a division of $L_{\rm bol}/L_{\rm Edd} \sim 0.01$ \citep{xu2009}.

It is usually suggested that there should be a critical accretion rate for ADAF transiting to SSD \citep{ghisellini2001,yuan2014}. 
\cite{cao2004} estimated the maximum jet power for a given black hole mass via the Blandford-Znajek mechanism \citep{blandford1977}, assuming a rapidly spinning BH, e.g., $a=0.95$. 
Then, they analyzed the optical and radio properties for a sample of 33 FR I radio galaxies selected by \cite{chiaberge1999}. It turns out that, on average, the jet power estimated from the observed radio luminosity at 151 MHz luminosity \citep{willott1999}, is one order of magnitude larger than the theoretical prediction for more than one-third of the sources (13 of 33 sources) in the sample. 
Note that there is no difference in the BH mass distribution between the 13 sources with more powerful jets and the rest. 
Thus, the presence of sources with higher jet power suggests that the accretion mode of FR I radio galaxies might be more complex than expected before \citep{ghisellini2001}.
Moreover, the optical core luminosities of sources with higher jet power are surprisingly of the same order of magnitude as the ones of the rest \citep{cao2004}. 
Therefore, the nature of those low-accretion-rate radio galaxies with powerful jets is still unclear, which is the motivation for this work.

In past studies, the powerful jets are also observed in other supermassive BHs with low accretion rates, e.g., radio-loud AGN, M87 and Sgr $\rm A^{\ast}$ \citep{zamaninasab2014,eht2021,eht2022,yuan2022}. 
\cite{igumenshchev2002} and \cite{igumenshchev2003} described 3D magnetohydrodynamic (MHD) simulations that poloidal and frozen-in magnetic flux is dragged to the center of BH by accretion, where it impedes the accreting gas, causing a great increase in the magnetic pressure and a substantial reduction in the gas velocity. The accumulated magnetic would disrupt the accretion flow, which is referred to as a magnetically arrested disk \citep[MAD;][]{narayan2003}. Recently, \cite{You2023Sci...381..961Y} discovered time lags of 8 and 17 days in radio and optical fluxes, compared to X-ray flux, by analyzing the outburst from the black hole X-ray binary MAXI J1820+070 in 2018. The radio delay of about 8 days, for the first time, reveals the formation of a MAD around the accreting BH. \cite{tchekhovskoy2011} demonstrated that the magnetically arrested disk might explain the observations of the powerful jets with the efficiency $\eta_{\rm jet} \approx$ few $\times100$ percent, which characterizes the ratio of the jet power to the mass-energy flowing into the BH, via the numerical simulations of accreting BHs. 
The jet power of radio-loud galaxies is greatly enhanced under the MAD model of the BH accretion \citep{zdziarski2015,lisl2022}. 

In this paper, we study a sample of FR I radio galaxies, analyzing their Eddington scaled accretion rates in relation to other observable properties in order to the nature of powerful jets driven by low accretion rates. In Section \ref{sec:data}, we present the black hole masses, the estimated jet power, the optical core luminosities, the X-ray luminosities, and the H$\beta$ luminosity for the sample. 
In Section \ref{sec:results}, the predicted jet powers are calculated via the Blandford-Znajek mechanism, with the low Eddington-scaled accretion rate estimated, and the calculation for the predicted jet powers with MAD is also described. In Section \ref{sec:discuss}, we discuss the further confirmation of the low accretion rate for the sample and the black hole spin for the sample.

The cosmological parameters $H_{0} = 70~\rm km~s^{-1}~Mpc^{-1}$ and $\Omega_0 = 0.3$ have been adopted in this paper.

\section{Sample and Data} \label{sec:data}
\cite{buttiglione2010,buttiglione2011} reported their own classification of the 3CR catalog with rather strict criteria based on the best radio maps in the literature. Only the sources with clear twin jets are considered as FR I radio galaxies, and only those with clear hot spots are included in FR II radio galaxies. This led to 21 FR I radio galaxies in total and a large number of undefined sources. By comparison with the sample in \cite{cao2004}, the BH masses of 17 of 21 sources were estimated using the relation between host galaxy absolute magnitude at R-band and black hole mass \citep{mclure2002}, i.e. $\log(M_{\rm BH}/{\rm M_{\odot}}) = -0.50(\pm0.02)M_{\rm R}-2.96(\pm0.48)$. Consequently, the sample consisting of 17 FR I radio galaxies is constructed in this work.

\begin{table*}
 \begin{center}
  \caption{Column (1): source name, Column (2): references for the identification of FR I, Column (3): redshift, Column (4): Logarithm of the jet power with the factor of $f=1$ from \protect\cite{cao2004}, Column (5): Logarithm of the optical core luminosity from \protect\cite{chiaberge1999}, Column (6): Logarithm of the H$\beta$ luminosity from \protect\cite{buttiglione2009,buttiglione2011}, Column (7): Logarithm of the X-ray luminosity in 2-10 keV, Column (8): references for the X-ray data, Column (9) Logarithm of the BH mass from \protect\cite{cao2004}. Notes: $\rm a$: The sources for which F791W has been used; $\rm b$: The sources for which F814W has been used; $\rm c$: The source for which F555W has been used; $\rm d$: The sources whose optical core luminosities are derived without decomposition of nuclear and galaxy; $\rm e$: The X-ray data in 2-7 keV band instead; $\rm f$: The X-ray data in 4.5-12 keV band instead. References: B10: \protect\cite{buttiglione2010}, B11: \protect\cite{buttiglione2011}, E20: \protect\cite{chandra2020}, CH17: \protect\cite{caglar2017}, M17: \protect\cite{moss2017}, G09: \protect\cite{gonzalez2009}.}
\label{table1}
\begin{tabular}{l |c|c  |c  |c|cccc }\hline\hline
Name &Ref.& Redshift & $\log Q_{\rm jet}~({\rm erg~s^{-1}})$&$\log L_{\rm c}~({\rm erg~s^{-1}})$&$\log L_{\rm H\beta}~({\rm erg~s^{-1}})$&$\log L_{\rm X}~({\rm erg~s^{-1}})$&Ref.& $\log M_{\rm BH}/\rm M_{\odot}$\\
 (1) & (2) & (3)&(4)&(5) &(6) &(7)& (8)&(9) \\ \hline
 3C~29 & B10 & 0.045 & 43 & 41.28 & 39.44 & 39.50$\rm ^e$ & E20 & 9.1\\
 3C~31 & B10 & 0.0169 & 42.3 & 40.82 & 39.01 & 39.63$\rm ^e$ & E20 & 8.6\\
 3C~66B & B10 & 0.0213 & 42.8 & 41.55 & 39.45 & 40.17$\rm ^e$ & E20 & 8.5\\
 3C~75$\rm ^d$ & B10 & 0.0232 & 42.7 & $<$41.92 & $<$39.92 & 41.11 & CH17 & 9.0\\
 3C~76.1$\rm ^{cd}$ & B10 & 0.0325 & 42.6 & $<$40.73 & $<$39.82 & 40.80$\rm ^e$ & E20 & 8.2\\
 3C~78 & B10 & 0.0287 & 42.7 & 42.49 & 38.99 & 41.46$\rm ^e$ & E20 & 8.9\\
 3C~83.1 & B10 & 0.0251 & 42.8 & 40.14 & $<$39.26 & 39.20$\rm ^e$ & E20 & 8.8\\
 3C~89 & B10 & 0.1386 & 44 & $<$40.85 & $<$40.55 & 41.94$\rm ^e$ & E20 & 8.6\\
 3C~264$\rm ^{a}$ & B10 & 0.0217 & 42.6 & 42.04 & 39.11 & 42.09$\rm ^f$ & M17 & 8.3\\
 3C~270$\rm ^{a}$ & B11 & 0.0075 & 42.1 & 39.76 & 38.56 & 40.73 & G09 & 8.6\\
 3C~272.1$\rm ^{b}$ & B10 & 0.00334 & 41.1 & 40.13 & 37.92 & 39.4 & G09 & 8.2\\
 3C~274$\rm ^{b}$ & B10 & 0.0044 & 42.8 & 41.19 & 38.73 & 40.84 & G09 & 9.5\\
 3C~296 & B10 & 0.0237 & 42.5 & 40.48 & 39.35 & 39.78$\rm ^e$ & E20 & 8.8\\
 3C~338 & B10 & 0.0304 & 43.2 & 41.17 & 39.51 & 40.41$\rm ^e$ & E20 & 9.1\\
 3C~438 & B10 & 0.29 & 44.8 & $<$41.87 & $<$41.28 & 42.60$\rm ^e$ & E20 & 8.6\\
 3C~449 & B10 & 0.0171 & 42.2 & 40.91 & 38.71 & 40.32$\rm ^e$ & E20 & 8.0\\
 3C~465 & B10 & 0.0302 & 43.1 & 41.44 & 39.38 & 40.66$\rm ^f$ & M17 & 8.6\\

\hline\hline
\end{tabular}
\end{center}
\end{table*}

The jet power can be estimated by several methods, including the radio lobe emission based on equipartition assumption \citep{willott1999}, the radio core-shift effect \citep{shabala2012}, spectrum modeling \citep{ghisellini2014}, the total energetic output divided by the age \citep{mahatma2019} and the X-ray cavity measuring \citep{birzan2008}. 
Here, \cite{cao2004} also estimated the jet power $Q_{\rm jet}$ for FR I radio galaxies in this sample by the way of \cite{willott1999} in approximation, with $Q_{\rm jet} \simeq 3\times 10^{38} f^{3/2} L_{151}^{6/7}~\rm W$, where $L_{151}$ is the total radio luminosity at 151 MHz in units of ${\rm 10^{28} ~ W ~ Hz^{-1} ~ sr^{-1}}$, although the radio luminosity is not so correlated with the jet power based on the optical core luminosity for some FR I sources.
The factor $f$ is the uncertainty that absorbs all the unknown factors, including the unknown magnetic field, age, the environment of the source, and the unknown ratio of non-radiating (e.g. protons) to radiating particles (e.g. electrons and/or positrons), which is usually assumed to be zero and introduces substantial scatter \citep{willott1999,croston2018}. The value of $f$ is suggested to be in the range of $1\le f \le 20$. 
For FR I radio galaxies, e.g., M87, $f=1$ is favored \citep{cao2004}, which is consistent with the previous work \citep{bicknell1999,young2002}. In this work, the jet power of the sample is also estimated by assuming $f=1$ in approximation, which is listed in Table \ref{table1}. 

Most of the sources in this sample have been observed with the Hubble Space Telescope \citep{chiaberge1999}. The HST observations were taken using the Wide Field and Planetary Camera 2 (WFPC2). The pixel size of the Planetary Camera, in which the target is always located, is $0.0455''$, and the $800\times800$ pixels cover a field of view of $36''\times36''$. \cite{chiaberge1999} processed the data of HST observations in the public archive for 16 of all 17 sources (3C 76.1 has not been observed), almost all sources in the sample are observed using the F702W filter, except two sources for which F791W and F814W are used, as part of the HST snapshot survey of 3C radio galaxies. We determined the optical core luminosities $L_{\rm c}$ in the observer frame with the flux data from \cite{chiaberge1999}. For 3C 75 with the bright, compact knot but covered by dust lanes,  the optical core luminosity is simply estimated without separation for nuclear and host galaxy by photometric aperture analysis for the entire galaxy, as the upper limit, and we make the same approximation for 3C 76.1 with F555W filter. 

\cite{buttiglione2009,buttiglione2011} presented the optical spectra for all 17 radio galaxies with the Telescopio Nazionale Galileo (TNG). The observations were made using the DOLORES (Device Optimized for the LOw RESolution) spectrograph installed at the Nasmyth B focus of the telescope. The measurements of the emission line intensities were done by fitting Gaussian profiles to H$\beta$, [O III]$\lambda$5007, [O I]$\lambda \lambda$6300,64, H$\alpha$, [N II]$\lambda \lambda$6548,84 and [S II]$\lambda \lambda$6716,31 with the {\it specfit} package in IRAF. The H$\beta$ luminosity and the main emission line ratios are shown in Table \ref{table1} and Table \ref{table2}, respectively.

We also search for the X-ray luminosity in the 2-10 keV band in the literature to estimate the bolometric luminosity in Sec \ref{sec:rate2}. However, the X-ray data in 2-10 keV band is available only for 3C 75, 3C 270, 3C 272.1 and 3C 274 \citep{caglar2017,gonzalez2009}. \cite{moss2017} reported the X-ray flux in the 4.5-12 keV band for 94 radio AGN, which contains 3C 264 and 3C 465 in our sample. For the rest of sources, the data in 2-7 keV band is available in {\it Chandra} catalogue \citep{chandra2020}. Table \ref{table1} lists the X-ray luminosity and references.

\section{Results} 
\label{sec:results}

\subsection{Eddington-scaled accretion rates}
\label{sec:rate}

The accretion mode can be inferred from the Eddington-scaled accretion rates $\dot{m}=\dfrac{\dot{M}}{\dot{M}_{\rm Edd}}$, where $\dot{M}_{\rm Edd}=\dfrac{L_{\rm Edd}}{\eta_{\rm eff}c^2}$ and $L_{\rm Edd}$ is the Eddington luminosity. Given a constant accretion efficiency, like the standard value of 0.1, the accretion rate can be estimated by the Eddington ratio, 
defined by $\lambda = L_{\rm bol}/L_{\rm Edd}$, where $L_{\rm bol}$ is the bolometric luminosity. 

\begin{figure}
    \centering
    \includegraphics[width=0.4\textwidth]{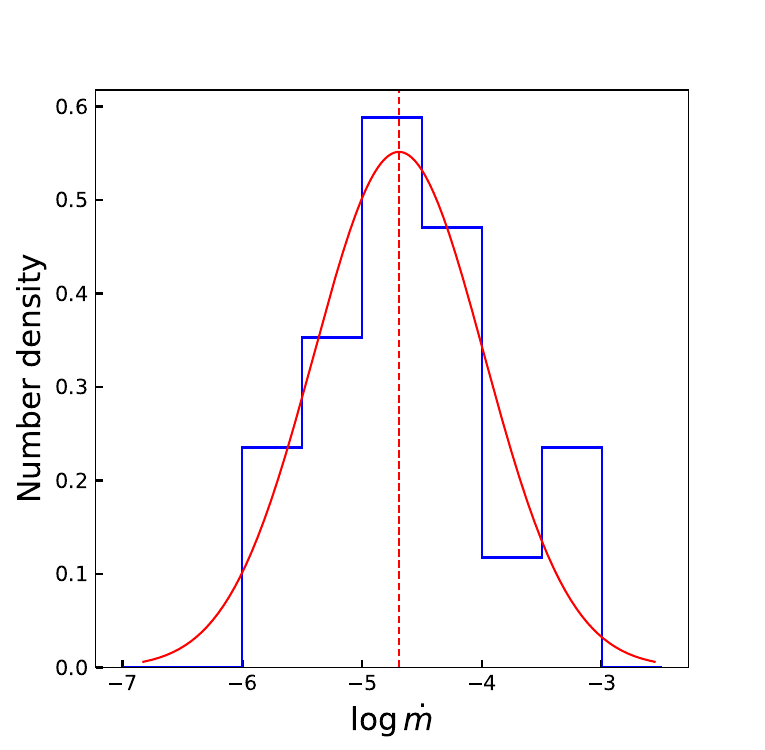}
    \caption{The accretion rate of our sample with binsize=0.5 from -7.0 to -2.5. The number density represents the ratio of the number of sources in each bin to the sample size. The red line represents the Gaussian distribution for the sample. The pink dashed line represents the median accretion rate.}
    \label{fig:rate}
\end{figure}

\cite{mclure2004} derived the calibration between absolute B-band magnitude and the bolometric luminosities for a sample of 372 sources in 2dF 10K quasar catalog \citep{croom2001} and the new SDSS Quasar Catalogue II \citep{schneider2003}. The relation is given by:
\begin{equation}
\label{eq:Mb&bol}
    M_{\rm B} = -2.66(\pm 0.05)\log [L_{\rm bol}/\rm W]+79.36(\pm 1.98).
\end{equation}
Based on this, the relation between optical B-band luminosity and bolometric luminosity can be derived by \citep{Kaspi2000ApJ...533..631K,wu2013}:
\begin{equation}
    \label{eq:Lbol}
    L_{\rm bol}~({\rm erg~s^{-1}}) = 10L_{\rm B},
\end{equation}
where $L_{\rm B}$ is the luminosity at 4400 \AA. Note that the central wavelengths in HST photometry are 6919 \AA, 7881 \AA, 8012\AA, and 5439 \AA, for F702W, F791W, F814W, and F555W, respectively. So we take the optical core luminosities $L_{\rm c}$ in Table \ref{table1} as $L_{\rm B}$. For the $L_{\rm c}$ with an upper limit, we take the upper limit into the calculation to obtain a larger accretion rate. As a result, the estimated bolometric luminosity in this way covers a range of $40.76<\log L_{\rm bol}<43.49$. With the derived BH mass in \cite{cao2004}, we find that the Eddington ratios of all the sources in the sample are substantially lower than 0.01, the value that separates low and high Eddington-scaled accretion, covering a range of $-6.0<\log \dot{m} <-3.4$ (see Fig. \ref{fig:rate}), which suggests that ADAFs are present in these sources. We assume the accretion rates for the sample satisfy the Gaussian distribution (see the pink line in Fig. \ref{fig:rate}), and find the median of $\mu=-4.7$ and the error of $\sigma=0.7$ in $\log \dot{m}$. 
In this case, we then estimate the maximal jet powers of the sample from an ADAF surrounding the BH with the mass $M_{\rm BH}$ and the spin $a$ \citep{livio1999}, in the Blandford-Znajek mechanism, by assuming the median accretion rate $\log \dot{m} = -4.7\pm0.7$.
The maximum estimated jet power extracted from a spinning black hole is given by: 
\begin{equation}
    L_{\rm BZ}=\left(\frac{B^2}{4\pi}\right) \pi R_{\rm h}^2 \left(\frac{R_{\rm h} \Omega_{\rm h}}{c} \right)^2c,
\label{eq:L_bz}
\end{equation}
where $R_{\rm h}$ is the radius of the horizon and $\Omega_{\rm h}$ is the angular velocity of BH. Both of them are the function of the black hole spin $a$ and the black hole mass $M_{\rm BH}$. Following the calculation in \cite{cao2004}, the magnetic pressure for ADAF, which is proportional to the accretion rate, satisfies the relation $p\sim B^2/8\pi$. Consequently, the jet power $L_{\rm BZ}$ is determined by the accretion rate $\dot{m}$, the spin $a$, and the black hole mass $M_{\rm BH}$.

\begin{figure}
    \centering
    \includegraphics[width=0.4\textwidth]{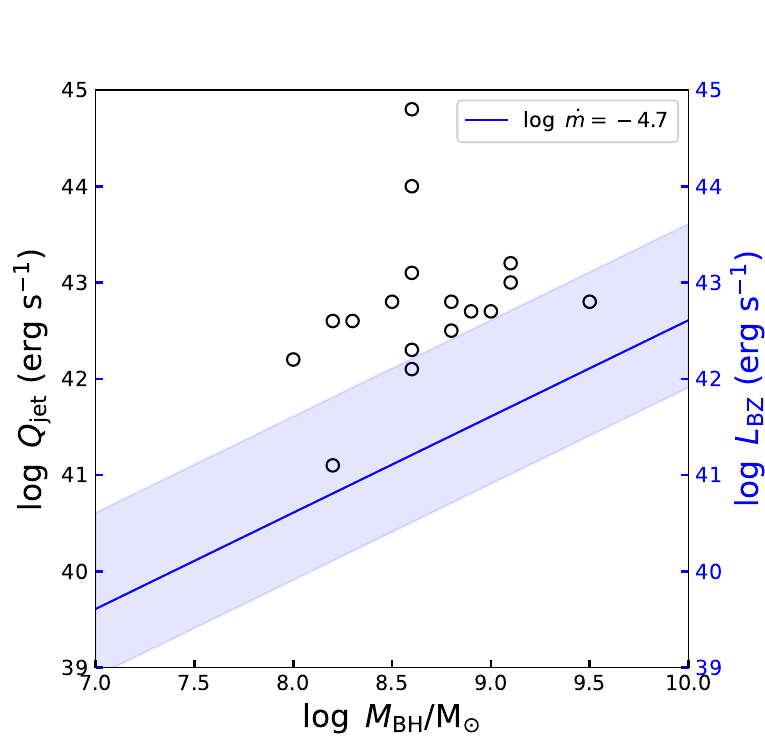}
    \caption{The BH mass $M_{\rm BH}$ versus the jet power $Q_{\rm jet}$. The empty circles correspond to the estimates from Table \ref{table1} for the sample. The blue line represents the predicted relation between the BH mass and the maximal jet power extracted from spinning black holes surrounded by ADAFs and the blue region represents the error for the median accretion rate $\log \dot{m}=-4.7\pm0.7$ in the sample. Note that the spin $a=0.95$ is used as the upper limit.}
    \label{fig:adaf_model}
\end{figure}

We plot the predicted relation between the jet power $Q_{\rm jet}$ and BH mass $M_{\rm BH}$ for BH spin $a =0.95$ in Fig. \ref{fig:adaf_model}. The estimated jet powers from the observed 151 MHz radio luminosities are also plotted. Surprisingly, the estimated jet powers $Q_{\rm jet}$ are greater than the maximum theoretical jet power for almost all the sources in the sample.
Apparently, the jet power is underestimated in the scenario of an ADAF. Note that, in the BZ mechanism, the jet is assumed to be driven by a spinning BH with a large-scale magnetic field, so the jet power depends on the field strength at the BH horizon \citep{livio1999}. Therefore, the deficit in the jet power mentioned above implies that the magnetic fields of the sources are underestimated, which will be solved below.

\subsection{High jet power for low accretion rate: MAD}
\label{sec:mad}

We have shown above that the Eddington ratios for most of the sources in this sample are smaller than $-3.4$ in logarithms. This suggests that the accretion flows in these sources are ADAFs. However, the jet power is underestimated in the scenario of an ADAF. Note that, in the BZ mechanism, the jet is assumed to be driven by a spinning BH with a large-scale magnetic field, so the jet power depends on the field strength at the BH horizon \citep{livio1999}. Therefore, the deficit in the jet power mentioned above implies that the magnetic fields of the sources are underestimated.

The powerful jets are also detected in other supermassive BHs with low accretion rates, e.g., radio-loud AGN, M87, and Sgr $\rm A^{\ast}$ \citep{zamaninasab2014,eht2021,eht2022,yuan2022}. The strong magnetic fields can be achieved by considering the accumulation of the magnetic field, i.e., MAD \citep{narayan2003, yuan2014}. It was assumed that the poloidal magnetic flux could be collected in the vicinity of the BH due to the cumulative accretion. As the magnetic flux continuously increases, the BH gravity acting on the accretion flow could be balanced by the magnetic pressure, leading to the field being prevented from escaping by the continued accretion as well as further enhancement of the magnetic flux. Therefore, in the following, we take the accumulation of the magnetic flux into account, to recalculate the jet power in the scenario of the MAD.

The accumulated magnetic field disrupts the accretion flow at a magnetospheric radius $R_{\rm m}$. The accretion flow within $R_{\rm m}$ forms a MAD and the magnetic field supports the gas against gravity \citep{narayan2003}, which is given by:
\begin{equation}
    B_{\rm MAD}\sim 1.5\times10^9(1-f_\Omega)^{1/2}\epsilon^{-1/2}M_{\rm BH}^{-1/2}\dot{m}^{1/2}R^{-5/4}~\rm G,
\label{eq:b_mad}
\end{equation}
where $\epsilon=v_{\rm R}/v_{\rm K}~\sim~0.01-0.1$, $v_{\rm R}$ is the velocity of gas in MAD and $v_{\rm K}$ is the free-fall velocity \citep{narayan2012,begelman2022,kawamura2022}. The factor $f_\Omega$ is the ratio of $\Omega$ and $\Omega_{\rm K}$, where $\Omega$ is the angular velocity of the accretion flow and $\Omega_{\rm K}$ is the free-fall angular velocity. Here we assume $\epsilon=0.01$, $f_{\Omega}=0.5$, $R \sim R_{\rm ISCO}$, where $R_{\rm ISCO}$ is the innermost stable circular orbit (ISCO), the value of which depends on the BH spin \citep{you2012} as:
\begin{equation}
    R_{\rm ISCO}=3 + z_2 - \left[ (3-z_1)  (3+z_1+2z_2) \right]^{1/2},
\label{eq:r_isco}
\end{equation}
where $z_1=1 + (1-a^2)^{1/3} [ (1+a)^{1/3} + (1-a)^{1/3} ]$, $z_2=(3a^2+z_1^2)^{1/2}$. Substituting Eq. \ref{eq:b_mad} and \ref{eq:r_isco}  into Eq. \ref{eq:L_bz}, the calculation for $L_{\rm BZ}$ will be:
\begin{equation}
    L_{\rm BZ}=5.625\times10^{17}(1-f_{\Omega})\epsilon^{-1}M_{\rm BH}^{-1}\dot{m}R_{\rm ISCO}^{-5/2}\frac{R_{\rm h}^4\Omega_{\rm h}^2}{c}.
\label{eq:final_Lbz}
\end{equation}
Therefore, given the BH mass, BH spin, and the accretion rate, the jet power in the scenario of the MAD can be predicted with Eq. (\ref{eq:final_Lbz}). 

\begin{figure}
    \centering
    \includegraphics[width=0.4\textwidth]{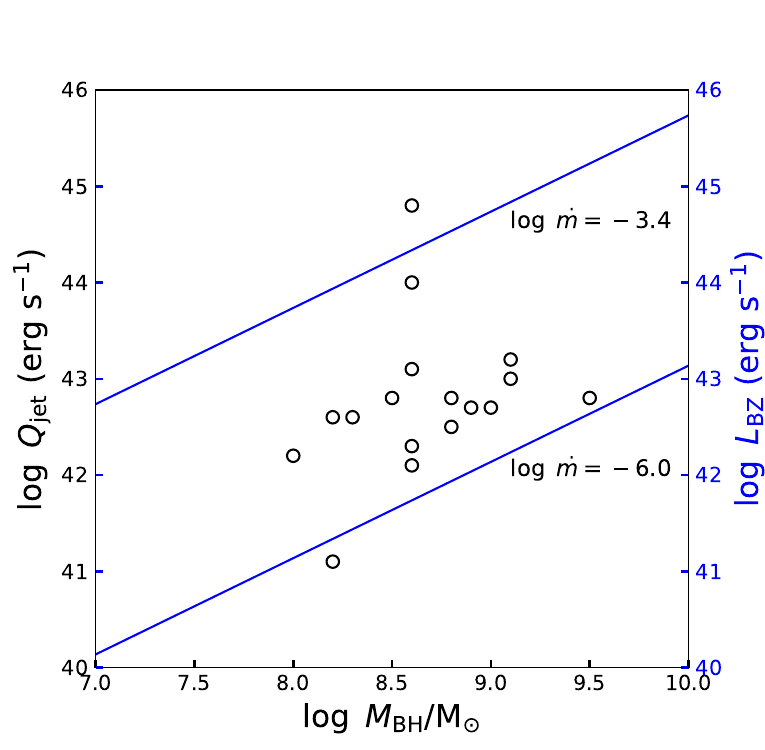}
    \caption{The BH mass $M_{\rm BH}$ versus jet power $Q_{\rm jet}$. The blue solid lines represent the predicted relation between the BH mass and the maximal jet power with Equation \ref{eq:final_Lbz}, in two cases: $\log \dot{m}=-6.0$ and $\log \dot{m}=-3.4$, by assuming $a=0.95$.}
    \label{fig:mad_0.95}
\end{figure}

In Sec. \ref{sec:rate} we have shown that the Eddington ratios which are estimated to be in the range of  -6.0 and -3.4 in logarithm, we predict the relation between the jet power and the BH mass with Equation (\ref{eq:final_Lbz}), in two cases: $\log \dot{m} = -6.0$ and $\log \dot{m} = -3.4$, by assuming $a=0.95$. It is found that the predicted jet power is now compatible with the estimated jet power (see Fig. \ref{fig:mad_0.95}), which indicates that the accretion flow is magnetically arrested so that the accumulation of the magnetic field is essential to efficiently power the jet.

\section{Discussion}
\label{sec:discuss}

In this work, we estimate the Eddington ratios to be from -6.0 to -3.4 in logarithms for the 17 FR I radio galaxies. These observational results indicate that the sample should be radiatively inefficient sources with low accretion rates. Given the low accretion rate, we demonstrate that the accumulation of the magnetic field in the MAD is required to account for the jet power in this sample.

\subsection{Other methods for accretion rate}
\label{sec:rate2}

\begin{figure*}
    \centering
    \includegraphics[width=1\textwidth]{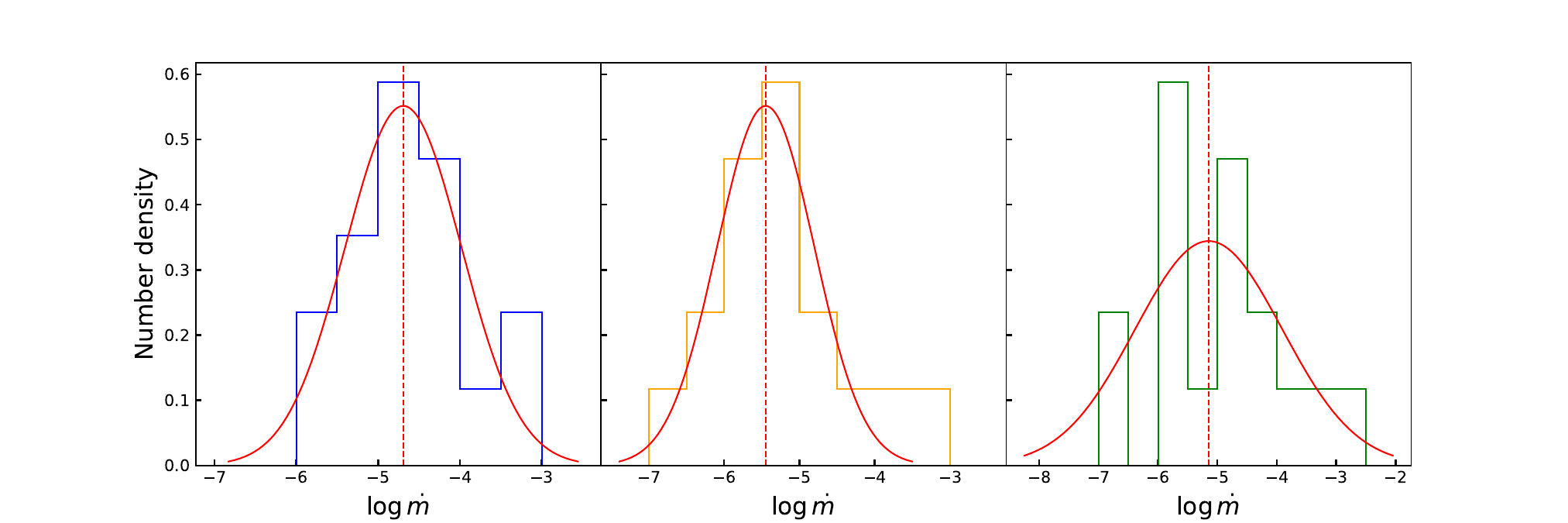}
    \caption{The accretion rate through three different methods with binszie=0.5 from -7.0 to -2.5. The number density represents the ratio of the number of sources in each bin to the sample size. The blue, red, and green line represents the relation between the bolometric luminosity $L_{\rm bol}$ and $L_{\rm B}$, $L_{\rm H\beta}$ and $L_{\rm X}$, respectively. The red lines and red dashed lines represent each method's Gaussian distribution and median accretion rate.}
    \label{fig:rate2}
\end{figure*}

We also estimated the bolometric luminosity through two different methods to have a thorough view of the accretion rate. The first is similar to the method in Sec. \ref{sec:rate}, but where the $M_{\rm B}$ is estimated by the relation between the nuclear H$\beta$ luminosity and absolute B-band magnitude \citep{ho&peng2001}, which is given by:
\begin{equation}
    \label{eq:Hb&Mb}
    \log L_{\rm H\beta} = (-0.34 \pm 0.012)M_{\rm B}(L_{\rm bol}) + (35.1\pm0.25),
\end{equation} where the H$\beta$  luminosities for the samples are presented in \cite{buttiglione2009,buttiglione2011}. Combine the Eq. \ref{eq:Hb&Mb} with the Eq. \ref{eq:Mb&bol}, the bolometric luminosity $L_{\rm bol,H\beta}$ covers a wide range of  $39.95<\log L_{\rm bol, H\beta}<43.67$. Similarly, with the derived BH mass in \cite{cao2004}, we find that the Eddington ratios of all the sources in the sample are substantially smaller than 0.01, the value that separates low and high Eddington-scaled accretion, covering a range of $-6.8<\log \dot{m} <-3.0$ (see Fig. \ref{fig:rate2}), with the median of $\mu =-5.4$ and the error of $\sigma=0.6$ in $\log \dot{m}$, by assuming a Gaussian probability density function. 
Note that the host galaxy contamination in the emission line is positive to our results: if the real nuclear H$\beta$ luminosity is smaller than we assume, the absolute B-band magnitude will be greater, and the bolometric luminosity will be smaller, leading to a lower range of accretion rates.

The second method is widely known through a luminosity-dependent bolometric correction in \cite{duras2020}. They analyzed nearly 1000 type 1 and type 2 AGNs and computed the bolometric correction separately for each group. They found that the two bolometric corrections are similar in the overlapping luminosity range and presented a general bolometric correction $K_{\rm X}$ from the X-ray luminosity (2-10 keV) $L_{\rm X}$ for both type 1 and type 2 AGNs, which gives:
\begin{equation}
    K_{\rm X} = a\left[1+\left(\frac{\log (L_{\rm X} / L_{\odot})}{b} \right)^c \right],
\end{equation}
where $a=15.33\pm0.06$, $b=11.48 \pm 0.01$, $c=16.20\pm0.16$. Then the bolometric luminosity will be:
\begin{equation}
    L_{\rm bol,X}~({\rm erg~s^{-1}}) = K_{\rm X}L_{\rm X}.
\end{equation}
Here, we take the X-ray luminosity in 2-7 keV band for most sources and 4.5-12 keV band for 3C 264 and 3C 465 into the calculation. In this case, the bolometric luminosity covers a range of $40.39~{\rm erg~s^{-1}}<\log L_{\rm bol, X}<43.79~{\rm erg~s^{-1}}$ and the accretion rates of the sample are in the range of $-6.5 < \log \dot{m} < -2.9$ (see Fig. \ref{fig:rate2}), with the median of $\mu=-5.1$ and the error of $\sigma=1.2$ in $\log \dot{m}$, by assuming a Gaussian probability density function.

It should be pointed out that, the accretion rates, through the latter two methods, lie in a wider range than the former, and they are consistent with results from $L_{\rm c}$. The medians for the latter two methods are smaller than the first one ($\mu=-4.7,-5.4,-5.1$, respectively), meaning the estimated accretion rates are truly small and credible for the sample.

\subsection{Baldwin–Phillips-Terlevich Diagram}
\label{sec:bpt}

In Sec. \ref{sec:rate} and \ref{sec:rate2}, we have shown that the sources tend to be of low accretion rates. We note that the Baldwin–Phillips-Terlevich (BPT) optical diagnostic diagrams, based on four emission line ratios [O III]/H$\beta$, [N II]/H$\alpha$, [S II]/H$\alpha$, [O I]/H$\alpha$, are also used in the literature to classify AGNs and probe the accretion mode \citep{baldwin1981,kewley2001,kauffmann2003,ho1997,kewley2006}. With these diagnostic diagrams, AGNs can be classified into at least two branches, i.e., Seyfert galaxies, and Low-Ionization Nuclear Emission-line Regions (LINERs). The Eddington ratios for LINERs are systematically smaller than 0.01 and are larger than 0.01 for Seyferts galaxies \citep{heckman2014}. \cite{kewley2006} also found that the transition between Seyferts and LINERs corresponds to an Eddington ratio of 0.05.

To intuitively distinguish the two branches through the BPT diagrams, \cite{ho1997} proposed a classification scheme for different classes using the nuclear emission-line ratios of 418 galaxies. They demonstrated that Seyfert galaxies are often defined to have [O III]/H$\beta \ge 3$, [N II]/H$\alpha \ge 0.6$, [S II]/H$\alpha \ge 0.4$ and [O I]/H$\alpha \ge 0.08$, LINERs to have [O III]/H$\beta < 3$, [N II]/H$\alpha \ge 0.6$, [S II]/H$\alpha \ge 0.4$ and [O I]/H$\alpha \ge 0.17$. Subsequently, \cite{kewley2006} selected a sample of $\sim$ 85 000 emission line galaxies from the SDSS, finding that Seyferts and LINERs form the separated branches on the diagnostic diagrams. They suggested that the dichotomy lines between Seyferts and LINERs are defined as:
\begin{equation}
    1.89\log \left({\rm \left[S~II \right]/H}\alpha \right)+0.76=\log({\rm [O~III]/H}\beta),
\end{equation}
\begin{equation}
    1.18\log({\rm [O~I]/H}\alpha)+1.3=\log({\rm [O~III]/H}\beta),
\end{equation}
where Seyfert galaxies lie above and LINERs lie below the dichotomy lines.

\begin{figure*}
    \centering
    \includegraphics[width=1\textwidth]{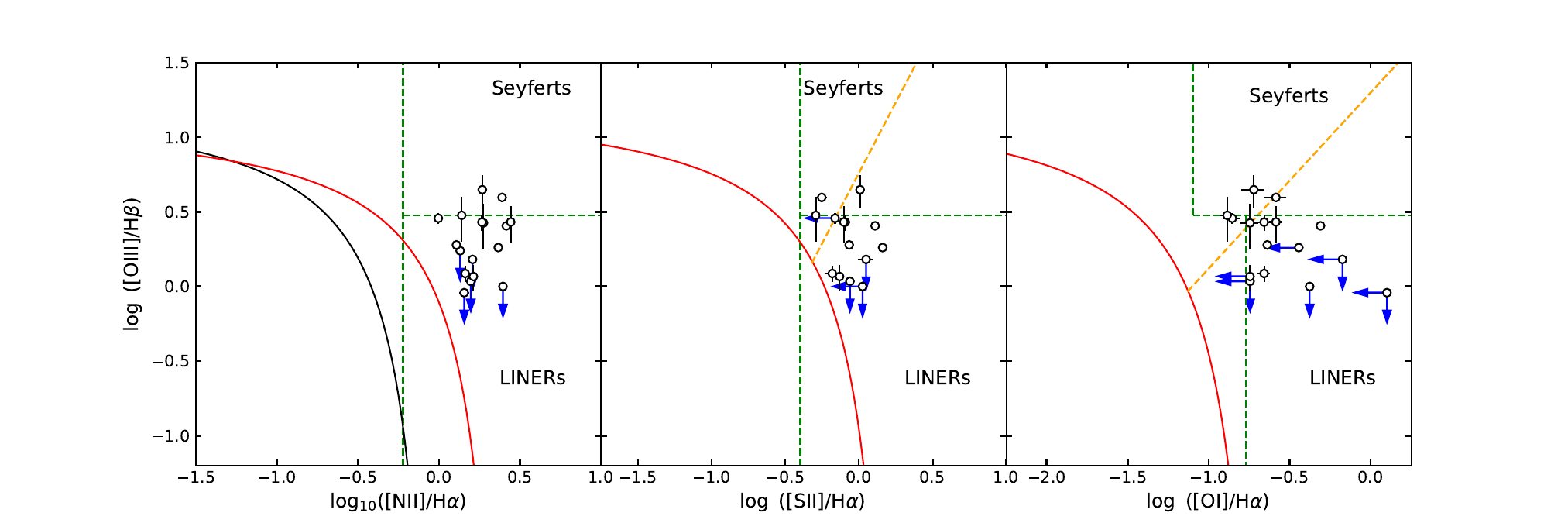}
    \caption{The three BPT diagrams for our sample. The red solid curve shows the division between starburst galaxies and AGNs defined by \protect\cite{kewley2001}. The black solid curve shows the revision in \protect\cite{kauffmann2003}. The green and orange dashed lines show the Seyfert-LINER classification schemes in \protect\cite{ho1997} and \protect\cite{kewley2006} respectively. The blue arrows represent the upper limit for the points.}
    \label{fig:BPT}
\end{figure*}

We plot the BPT diagram for the sample in Fig. \ref{fig:BPT}, based on the three classification schemes mentioned above. Table \ref{table2} presents the main emission line ratios. It is found that the majority of the sources in our sample belong to AGNs with high ratios of [O III]/H$\beta$ and low ratios of [N II]/H$\alpha$ and [SII]/H$\alpha$. More specifically, excluding the sources without some emission line data, 88\% of sample in [N II]/H$\alpha$ -- [O III]/H$\beta$ diagram, 86\% of the sample in [S II]/H$\alpha$ -- [O III]/H$\beta$ diagram, 75\% of sample in [O I]/H$\alpha$ -- [O III]/H$\beta$ diagram, belongs to LINERs, according to the classification scheme of \cite{ho1997}.
In addition, 86\% of the sample in [S II]/H$\alpha$ versus [O III]/H$\beta$ diagram, 81\% of the sample in [O I]/H$\alpha$ versus [O III]/H$\beta$ diagram are classed as LINERs, according to the classification scheme of \cite{kewley2006}. Therefore, the AGNs' demography above suggests that most of the sample may be `radiatively inefficient' or `low-excitation' LINERs, which is consistent with the conclusions of low accretion rates in Sec. \ref{sec:rate}.

\begin{table*}
 \begin{center}
  \caption{Column (1): source name, Column (2)-(5): the main emission line ratios from \protect\cite{buttiglione2009,buttiglione2011}. Note that the data for 3C 270 are taken from \protect\cite{ho1997} and no errors were reported.}
\label{table2}
\begin{tabular}{l |c c c c  |}\hline\hline
Name & [O III]$\lambda$5007/H$\beta$ & [N II]$\lambda$6584/H$\alpha$ & [O I]$\lambda$6364/H$\alpha$ & [S II]$\lambda \lambda$6716,31/H$\alpha$ \\ 
 (1) & (2) & (3) & (4) & (5) \\\hline 
 3C~29 & 4.46 $\pm$ 1.12 & 1.85 $\pm$ 0.02 & 0.19 $\pm$ 0.03 & 1.02 $\pm$ 0.04 \\
 3C~31 & 2.87 $\pm$ 0.24 & 0.99 $\pm$ 0.01 & 0.14 $\pm$ 0.02 & $<$0.69 \\
 3C~66B & 3.95 $\pm$ 0.28 & 2.45 $\pm$ 0.02 & 0.26 $\pm$ 0.04 & 0.56 $\pm$ 0.05 \\
 3C~75 & $<$1 & 2.48 $\pm$ 0.02 & 0.42 $\pm$ 0.01 & $<$1.06 \\
 3C~76.1 & $<$1.08 & 1.57 $\pm$ 0.02 & $<$0.18	$\pm$	0.00 & 0.87 $\pm$ 0.01 \\
 3C~78 & 2.67 $\pm$ 0.91 & 1.88 $\pm$ 0.04 & 0.18 $\pm$ 0.02 & - \\
 3C~83.1 & $<$1.74 & 1.35 $\pm$ 0.04 & - & - \\
 3C~89 & $<$0.91 & 1.43 $\pm$ 0.10 & $<$1.26 & - \\
 3C~264 & 1.22 $\pm$ 0.15 & 1.45 $\pm$ 0.01 & 0.22 $\pm$ 0.02 & 0.66 $\pm$ 0.08 \\
 3C~270 & 2.55 & 2.60 & 0.49 & 1.29 \\
 3C~272.1 &	1.90 $\pm$ 0.10 & 1.28 $\pm$ 0.01 & 0.23 $\pm$ 0.01 & 0.86 $\pm$ 0.01 \\
 3C~274 & 1.82 $\pm$ 0.03 & 2.32 $\pm$ 0.02 & $<$0.36 & 1.45 $\pm$ 0.01 \\
 3C~296 & 2.70 $\pm$ 0.33 & 1.84 $\pm$ 0.02 & 0.22 $\pm$ 0.05 & 0.81 $\pm$ 0.04 \\
 3C~338 & 1.17 $\pm$ 0.23 & 1.63 $\pm$ 0.02 & $<$0.18 & 0.74 $\pm$ 0.01 \\
 3C~438 & $<$1.52 & 1.61 $\pm$ 0.02 & $<$0.67 & 1.12 $\pm$ 0.13 \\
 3C~449 & 3.00 $\pm$ 1.00 & 1.38 $\pm$ 0.01 & 0.13 $\pm$ 0.01 & 0.51 $\pm$ 0.02 \\
 3C~465 & 2.71 $\pm$ 0.75 & 2.77 $\pm$ 0.03 & 0.26 $\pm$ 0.02 & 0.79 $\pm$ 0.01 \\

\hline\hline
\end{tabular}
\end{center}
\end{table*}

\subsection{Estimates for black hole spins}
\label{sec:bpt}

\begin{figure}
    \centering
    \includegraphics[width=0.4\textwidth]{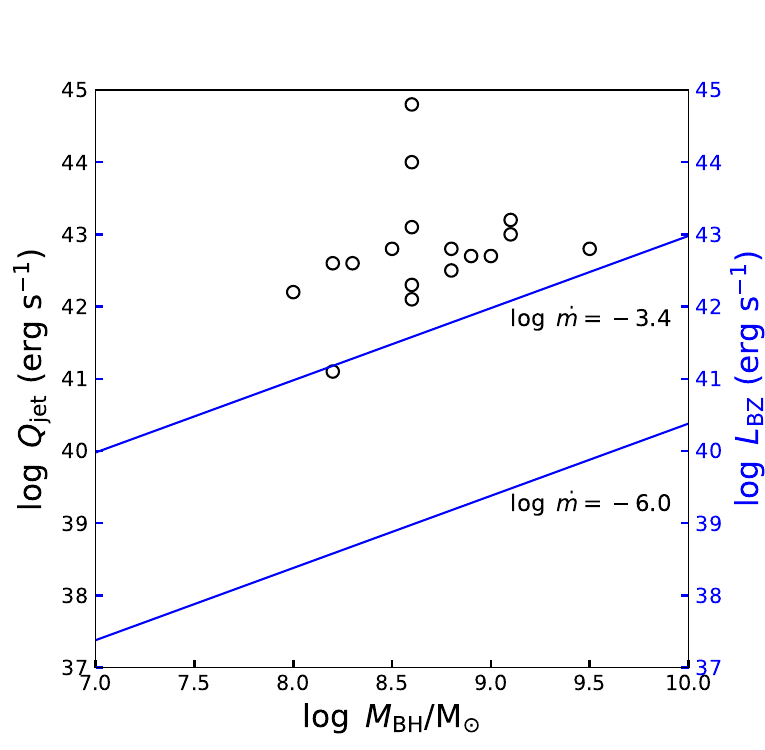}
    \caption{The BH mass $M_{\rm BH}$ versus jet power $Q_{\rm jet}$. The blue dashed lines represent the predicted relation between the BH mass and the maximal jet power with Equation \ref{eq:final_Lbz}, in two cases: $\log \dot{m}=-6.0$ and $\log \dot{m}=-3.4$, by assuming $a=0.1$.}
    \label{fig:mad_0.1}
\end{figure}

It should be noted that the jet power also depends on BH spin \citep{livio1999,narayan2003}. In the model calculations above (Fig. \ref{fig:mad_0.95}), fast-spinning BHs are assumed with $a=0.95$. Here, we can estimate the maximal jet power in the case of slow-spinning BHs with $a=0.1$. The relation between the jet power and BH mass for LJP sources, we can also estimate the maximal jet power in the case of slow-spinning BHs with $a=0.1$. The relation between the jet power and BH mass for LJPs is plotted in Fig. \ref{fig:mad_0.1}, assuming $\log \dot{m}=-3.4$ and $\log \dot{m}=-6.0$ as the upper and lower limit, respectively. It shows that the majority of sources lie above the predicted relation, which indicates that the predicted maximal jet power is smaller than the estimated jet power $Q_{\rm jet}$, even in the scenario MAD. This suggests that BHs in LJPs are Kerr BHs with $a>0.1$.

\begin{figure}
    \centering
    \includegraphics[width=0.4\textwidth]{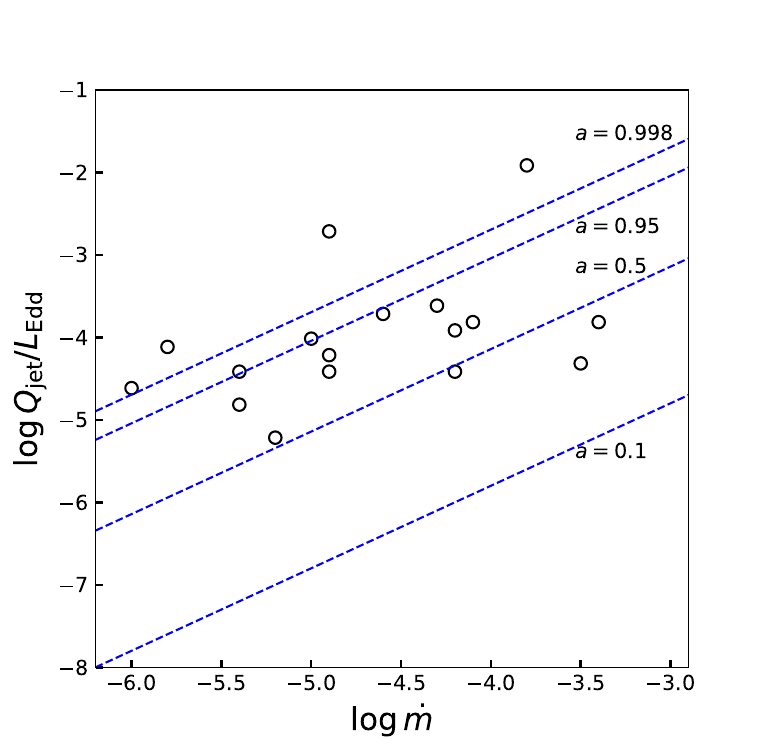}
    \caption{The relation between the accretion rate and the ratio of the jet power and the Eddington luminosity. The dashed lines represent the predicted relation between the accretion rate and the ratio of the maximal jet power with Equation \ref{eq:final_Lbz} and the Eddington luminosity, for different BH spins.}
    \label{fig:rate_Q_Ledd}
\end{figure}

Moreover, it is clear that the maximal estimated jet power is proportional to the BH mass if substituting $R_{\rm h}=\dfrac{GM_{\rm BH}}{c^2}\left[1+\left(1-a^2 \right)^{1/2}\right]$ and $\Omega_{\rm h}=\dfrac{ca}{2R_{\rm h}}$ into Equation (\ref{eq:final_Lbz}), which gives:
\begin{equation}
    L_{\rm BZ}={\rm C}M_{\rm BH}\dot{m}R_{\rm ISCO}^{-5/2}a^2 \left[1+\left(1-a^2 \right)^{1/2} \right]^2,
\end{equation}
where ${\rm C} \simeq 1.41\times10^{17}(1-f_{\Omega})\epsilon^{-1}G^2c^{-3}$. Therefore, we plot the relationship between the accretion rate and the ratio of the jet power and Eddington luminosity in Fig. \ref{fig:rate_Q_Ledd}, for $a=0.998, 0.95, 0.5$, and 0.1. It is found that the majority of sources lie above the predicted line for $a=0.5$, which suggests that the BHs in the sample are likely modest/fast-spinning with $a\gtrsim0.5$.

\subsection{Potential effect of the undefined sources}
It is worth noting that 21 undefined sources in \cite{buttiglione2010} are not included in the sample of FR I galaxies in our study. And only 10 of 21 undefined sources had the BH masses measurement and HST optical observations with F702W filter \citep{cao2004,chiaberge1999}, which are listed in Table \ref{tab:table3}. The optical data for 3C 293, 3C 315, and 3C 433 were not processed due to the dust lanes covering the central regions \citep{chiaberge1999}. Similarly, we estimate their optical core luminosity by photometric aperture analysis for the entire galaxy, without separation for nuclear and host galaxy.

\begin{figure}
    \centering
    \includegraphics[width=0.4\textwidth]{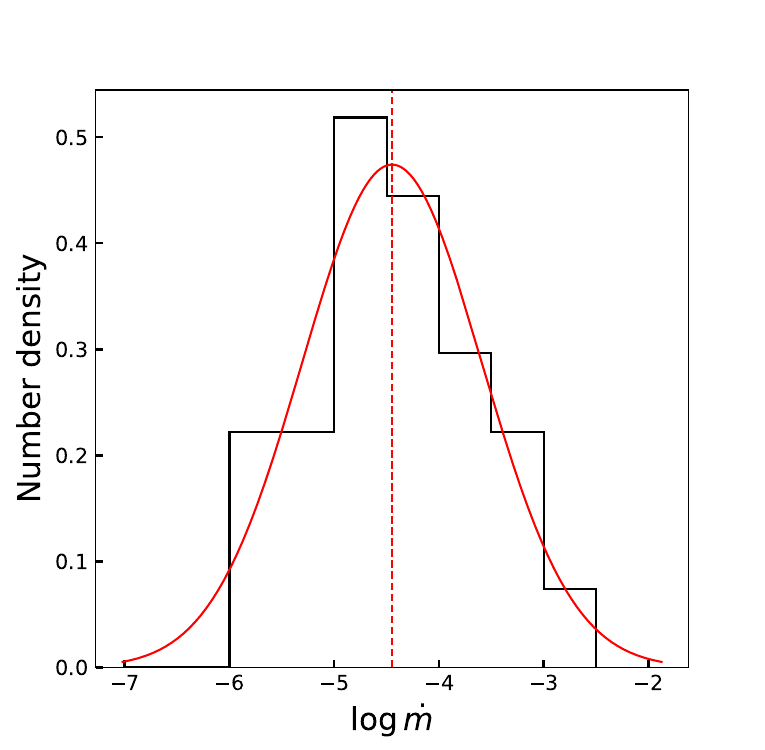}
    \caption{The accretion rate of the sample where the undefined sources are included. The bins are from -7.0 to -2.5 with binsize=0.5. The number density represents the ratio of the number of sources in each bin to the sample size. The red line represents the Gaussian distribution for the sample. The red dashed line represents the median accretion rate.}
    \label{fig:un_rate}
\end{figure}

\begin{figure}
    \centering
    \includegraphics[width=0.4\textwidth]{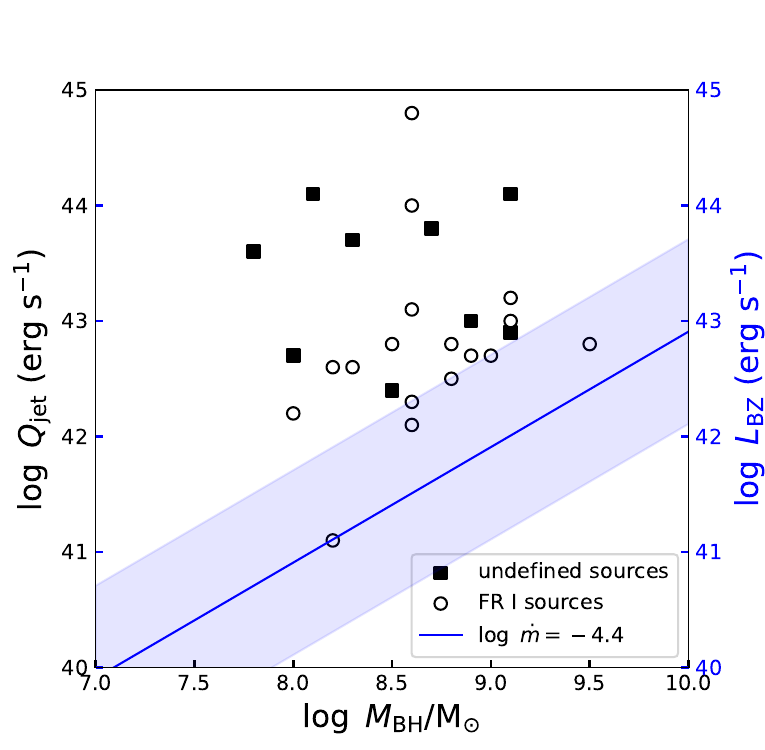}
    \caption{The BH mass $M_{\rm BH}$ versus the jet power $Q_{\rm jet}$. The empty circles correspond to Table \ref{table1} estimates for the FR I sample. The black squares correspond to the estimates from Table \ref{tab:table3} for the undefined sources. The blue line represents the predicted relation between the BH mass and the maximal jet power extracted from spinning black holes surrounded by ADAFs and the blue region represents the error for the median accretion rate $\log \dot{m}=-4.4\pm0.8$ in the sample. Note that the spin $a=0.95$ is used as the upper limit.}
    \label{fig:un_adaf}
\end{figure}

In this section, we include 10 undefined sources in our FR I sample, to discuss how these undefined sources might affect the results. Similarly, we take $L_{\rm c}$ as $L_{\rm B}$ and estimate the accretion rate using Equation \ref{eq:Lbol}. This results in a range of $-6.0<\log \dot{m}<-2.9$ with the median of $\mu=-4.4$ and the errors of $\sigma=0.8$, by assuming a Gaussian probability density function, which is plotted in Fig. \ref{fig:un_rate}. Given the accretion rate, we calculate the maximal jet power in the scenario of the ADAF surrounding the BH, with the median accretion rate $\log \dot{m}=-4.4\pm0.8$ adopted and $a=0.95$ assumed. The results are plotted in Fig. \ref{fig:un_adaf}. It turns out the underestimated theoretical jet power for the majority of the sources. Additionally, in the scenario of the MAD, we predict the relation between the jet power and BH mass with Equation \ref{eq:final_Lbz} in two cases: $\log \dot{m}=-5.1$ and $\log\dot{m}=-2.8$, by assuming $a=0.95$. It is also found the predicted jet power in the scenario of the MAD is compatible with the jet power estimated from the radio luminosity. As a result, our main conclusions, e.g. the great jet power can be achieved only when the accretion flow is magnetically arrested, are still valid even including these 10 undefined sources as the candidates in the FR I samples.

\begin{table*}
    \centering
    \caption{Column (1): source name, Column (2): references for identification of FR I, Column (3): redshift, Column (4): Logarithm of the jet power with the factor of $f=1$ from \protect\cite{cao2004}, Column (5): Logarithm of the optical core luminosity from \protect\cite{chiaberge1999}, Column (6): Logarithm of the BH mass from \protect\cite{cao2004}. Notes *: The sources whose optical core luminosities are derived without decomposition of nuclear and galaxy. Reference: B10: \protect\cite{buttiglione2010}.}
    \begin{tabular}{lccccc}\hline\hline
Name & Ref. & Redshift & $\log Q_{\rm jet}~(\rm erg~s^{-1})$ & $\log L_{\rm c}~(\rm erg~s^{-1})$ & $\log M_{\rm BH}/{\rm M_{\odot}}$ \\ 
(1) & (2) & (3) & (4) & (5) & (6)\\ \hline
3C~028 & B10 & 0.1952 & 44.1 & $<$41.38 & 8.1\\
3C~084 & B10 & 0.0176 & 42.9 & 42.86 & 9.1\\
3C~293$^*$ & B10 & 0.045 & 43.0 & $<$41.81 & 8.9\\
3C~314.1 & B10 & 0.1197 & 43.6 & $<$41.23 & 7.8\\
3C~315$^*$ & B10 & 0.1083 & 43.8 & $<$41.75 & 8.7\\
3C~348 & B10 & 0.154 & 45.1 & 41.55 & 8.9\\
3C~386 & B10 & 0.017 & 42.4 & 42.80 & 8.5\\
3C~424 & B10 & 0.127 & 43.7 & $<$41.64 & 8.3\\
3C~433$^*$ & B10 & 0.1016 & 44.1 & $<$41.72 & 9.1\\
3C~442 & B10 & 0.0263 & 42.7 & 40.00 & 8.0\\ \hline\hline

    \end{tabular}
    \label{tab:table3}
\end{table*}

\begin{figure}
    \centering
    \includegraphics[width=0.4\textwidth]{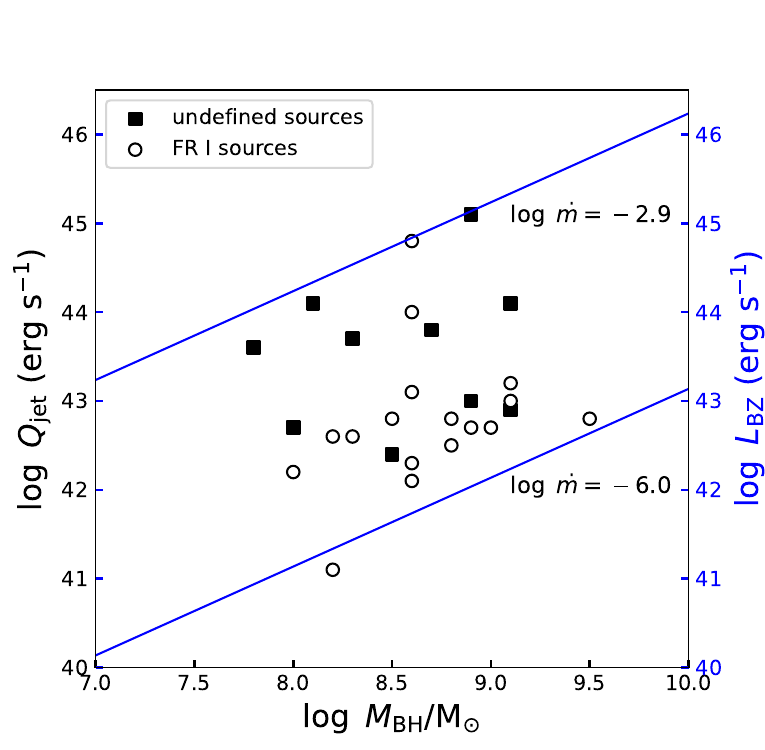}
    \caption{The BH mass $M_{\rm BH}$ versus jet power $Q_{\rm jet}$ for the undefined sources. The blue solid lines represent the predicted relation between the BH mass and the maximal jet power with Equation \ref{eq:final_Lbz}, in two cases: $\log \dot{m}=-6.0$ and $\log \dot{m}=-2.9$, by assuming $a=0.95$.}
    \label{fig:un_mad_0.95}
\end{figure}
\section{Summary}

We investigate the Eddington scaled accretion rates in relation to the observation properties for a sample of 17 FR I radio galaxies, to reveal the nature of the powerful jets with low accretion rates. The main results are as follows:

\begin{enumerate}
     \item The Eddington scaled accretion rates are estimated to be $\log \dot{m} \lesssim -3$, indicating the ADAFs are present in the sample, which is also confirmed by the BPT diagrams.

    \item The estimated jet power indicates a strong magnetic field within the ADAF, which suggests the existence of the MAD in our sample of FR I galaxies.

    \item 
    In the MAD scenario, modest/fast spins of $a\gtrsim0.5$ are favored for the BHs in this sample of FR I galaxies, by comparing the jet power and the Eddington luminosity for different accretion rates. 
\end{enumerate}

\section*{Acknowledgements}

We thank Heng-xiao Guo, Wenke Ren, Jiazheng Zhu, Min-Feng Gu, and Shuang-Liang Li for their helpful discussion/comments.
B.Y. is supported by NSFC grants 12322307, 12273026, and 12361131579; by the National Program on Key Research and Development Project 2021YFA0718500; by the Natural Science Foundation of Hubei Province of China 2022CFB167; by the Fundamental Research Funds for the Central Universities 2042022rc0002; Xiaomi Foundation / Xiaomi Young Talents Program.
X.C. is supported by the NSFC grants 12073023, 12233007, 12361131579,
and 12347103; by the science research grants from the China Manned Space Project with No. CMS-CSST- 2021-A06, and the fundamental research fund for Chinese central universities (Zhejiang University). J.-F. H is supported by the Projects funded by the Science and Technology Department of Qinghai Province (No. 2019-ZJ-A10). The Project is also funded by China Postdoctoral Science Foundation: 2020M682013.

\section*{Data Availability}
The optical core luminosities are available in \cite{chiaberge1999} and HST website (\url{https://mast.stsci.edu/search/ui/#/hst}). The H$\beta$ luminosity and emission line ratios come from \cite{buttiglione2009,buttiglione2011}. The X-ray data comes from NASA/IPAC Extragalactic Database (\url{http://ned.ipac.caltech.edu/}). The rest data can be obtained in \cite{cao2004}.



\bibliographystyle{mnras}
\bibliography{example} 





\bsp	
\label{lastpage}
\end{document}